\documentstyle[twocolumn,prc,aps,epsfig]{revtex}
\input epsf
\newcommand{\be}{\begin{eqnarray}}
\newcommand{\ee}{\end{eqnarray}}

\newcommand{\dd}{{\bf d}}
\newcommand{\dbar}{{\overline{\bf d}}}
\newcommand{\bb}{{\bf b}}
 \newcommand{\gsim}{\mathrel{\hbox{\rlap{\lower.55ex \hbox {$\sim$}}
                   \kern-.3em \raise.4ex \hbox{$>$}}}}
\newcommand{\lsim}{\mathrel{\hbox{\rlap{\lower.55ex \hbox {$\sim$}}
                   \kern-.3em \raise.4ex \hbox{$<$}}}}

\def\roughly#1{\mathrel{\raise.3ex\hbox{$#1$\kern-.75em%
\lower1ex\hbox{$\sim$}}}}
\def\lsim{\roughly<}
\def\gsim{\roughly>}

\begin{document}

\twocolumn[\hsize\textwidth\columnwidth\hsize\csname @twocolumnfalse\endcsname

\title {{Understanding the Non-Perturbative Deep-Inelastic Scattering:\\
 Instanton-induced Inelastic Dipole-Dipole Cross Section}}
\author {Edward V.Shuryak and Ismail Zahed}
\address { Department of Physics and Astronomy\\ State University of New York,
     Stony Brook, NY 11794-3800}

\date{\today}
\maketitle
\begin{abstract}
We derive the semiclassical (instanton-induced) contribution to the 
inelastic cross section of two color dipoles at large $\sqrt{s}$. We
study its dependence on the dipole sizes, orientations and most importantly the
impact parameter. The inelastic cross section is approximately quadratic in the
dipole sizes, and Gaussian-like in the impact parameter with a width of
the order of the instanton size. These results are directly relevant
to double DIS $\gamma^*\gamma^*$, as well as $\gamma^*\gamma$ and
standard DIS $\gamma^* h$ at small x when a real photon and a hadron 
can be approximated by a dipole. For such cases, with one small dipole
scattering on a large dipole, the impact parameter profile exhibits  a width
of about 1/2 fm, which is in good agreement with the impact
parameter profile recently extracted from DIS HERA data,
including diffractive $\gamma^*\rightarrow J/\psi$.   
\end{abstract}
\vspace{0.1in}
]
\begin{narrowtext}
\newpage

\section{Introduction}
\subsection{A puzzling  ``small gluonic spot'' }
Deep Inelastic Scattering (DIS) of leptons on a nucleon is
one of the best studied processes in high energy physics,
a benchmark for QCD applications since its early 
days. Still it continues to surprise us, with new data raising
yet new questions.

Perturbative  (pQCD) treatment of the $Q^2$ evolution, first by 
the OPE methods and moments, and then by 
the DGLAP evolution equation~\cite{DGLAP} is now a 
textbook material, correctly describing the $Q^2$ dependence at large 
enough $Q$. Surprisingly, the non-perturbative aspects of DIS 
were not much discussed in the literature, and in practice they were
always treated as a ``phenomenological input'' into the DGLAP evolution, 
usually at some low  scale $Q\approx 1 \, {\rm GeV}$. 

In the 1970's there was hope that all the glue in the nucleon is basically 
radiated from valence quarks, in a process described by familiar DGLAP
splitting functions. This description implies that the 3 valence quarks
in the nucleon should be used as inputs without any $\bar{q}q$ sea or
intrinsic glue. However, DIS data, especially from HERA  (see
e.g.\cite{DISdata}), do not support such views and some intrinsic
glue must be present even at the lowest normalization scale, both at
large and small x. The physical nature of this intrinsic glue is the
main issue to be discussed in this paper.

One important discovery made at HERA relates to diffractive DIS with a
surprisingly large cross section (about 10\% of the total cross
section). On general grounds one may think that a survival of the nucleon,
in spite of a violent DIS collision, means that the gluonic objects
hit in the process may be loosely related to the core of the
nucleon, or far from its valence quarks.

This discovery  also triggered a shift towards treatment of DIS in 
the nucleon rest frame, in which a virtual photon is substituted by a
dipole, frozen in size during its passage through the target. 
The empirical dependence on the dipole size and energy at low $x$ was
discussed by Golec-Biernat and Wusthoff~\cite{GBW}, Frankfurt,
McDermott and  Strikman~\cite{fms} and others. 
They all claimed that the dipole cross section shows a particular
scaling with dipole sizes, above which the cross section saturates.

The second important discovery came
from  the analysis of  diffractive DIS, such as 
$\gamma^*p\rightarrow J/\psi\,X\,p$. In particular, 
Kopeliovich and his collaborators~\cite{Kopel}
argued that gluon fluctuations from quarks cannot be described perturbatively
and should be restricted to {\it small spots} of the order of
$0.3$ fm in the trasverse direction.

In a paper devoted to global analysis of DIS
using non-linear evolution towards small $x$, Gotsman
et~al~\cite{Genya_etal} have observed that the input mean square
transverse size of the glue  $ R_\perp^2$
(at $x\approx 10^{-2}$), shows the best fit for its value to be about
3 GeV$^{-2}$, again much smaller than the electromagnetic radius.  

In a more recent paper devoted to the analysis of this  process
by Kowalski and Teaney~\cite{TK},  the dipole model
was generalized to local expressions in transverse plane. 
The impact parameter profile of the glue in the
nucleon was found to be small, with an rms width  much smaller 
than the nucleon  electromagnetic 
radius.  They also related the small size of the
gluonic spot to small nuclear shadowing for gluons.
They argued that shadowing is small simply because  
{\it small spots of glue} do not overlap in the impact parameter space.
A direct confirmation of such a picture came more recently from d-Au
collisions at RHIC. Instead of the widely expected saturation, it was found  that
even a nucleon going through the {\it diameter} of the Au nucleus, with
more than a dozen  nucleons on its way, experiences only small
shadowing corrections to its gluonic structure function that
are even overcome by (modest) Cronin effects.

Outside DIS, it is known that soft diffractive hadronic phenomena are 
related with small-size gluonic objects. For instance, the soft pomeron
parameters are related to the small gluonic objects, with the
pomeron slope $\alpha'\approx 1/(2\, {\rm GeV})^2\approx (0.1\, {\rm fm})^2$
setting the scale.  More specifically, it is well established
that the pomeron-related form-factors are {\it harder} than the 
electromagnetic ones. Those form-factors are completely consistent
with the profile $T(b)$ discussed in~\cite{TK}. In general,
the DIS data can be well parameterized by a two-pomeron model 
(soft plus hard)~\cite{DL}. Although at large $Q^2$ the DIS data
show some quantitative differences, we will assume that all diffractive
phenomena are of common origin.

In view of all these results, it appears that the current belief -- 
{\em that the gluons are radiated from the valence quarks in a cascade of
radiative  processes} -- should be revisited. Indeed, how is it 
possible that a  random radiation process starting with valence quarks
within a transverse size of the order of the electromagnetic radius,
result in a spot {\it smaller} in size than the electromagnetic
radius, the spread of the original valence quarks?

\begin{figure}[h]
\centering
\includegraphics[width=4cm]{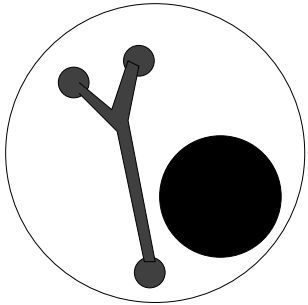}
\vskip 0.5cm
\caption{\label{fig_spot} 
A schematic view of a  nonperturbative glue distribution in a nucleon.
The black spot is an instanton.
}
\end{figure}

\subsection{The non-perturbative pomeron and DIS}

After outlining the phenomenological puzzle to be addressed,
let us move to its theoretical explanation. In short, we will argue
that the small-size gluonic object appearing both in diffractive
hadronic and DIS collisions at high energies are {\em instantons}
(leading to sphaleron production),
 coherent non-perturbative objects describing large-amplitude
fluctuations of the gauge fields in  the QCD vacuum. They are dominant
for small-$x$ gluons because their field strengths are larger
than that of the accompanying valence quarks. A schematic picture of
a nonperturbative glue in a nucleon, as seen by a passing dipole,
is shown in Fig.
\ref{fig_spot}.

But before we describe this in detail, we first comment on
previous important contributions to this field. The non-perturbative
approaches to this problem have been sought in the context of
the stochastic vacuum model by the Heidelberg group~\cite{SVM}.
In this model effective gluonic fields are reduced to Gaussian
fluctuations, with ``non-perturbative propagators''
used for gluon exchanges, and no non-linear terms or interaction
vertices. As a result, this model provides a successful account
of the constant cross section, but does not account for its
growth with energy, since no gluons or other objects can be 
promptly produced. In this sense, these works  focus  on the
description of the constant term $\sigma_0$ in the hh cross 
section

\be \label{eqn_crosssect}
\sigma(s)=\sigma_0 s^\Delta \approx \sigma_0 +\sigma_1 \,{\rm ln}\, s+...
\label{00}
\ee
Our discussion below is focused on the $\sigma_1$ term. The logarithmic
dependence on  the energy comes from the longitudinal phase space for 
the production of QCD sphalerons, which follow the instanton
excitation. In this sense, our approach and the one pursued by the 
Heidelberg group are complementary.

The relevance of instantons to DIS at large $x$ was discussed a decade ago by
Balitsky and  Braun \cite{BB_93} and also Ringwald and
Schrempp~\cite{RS}. The 
latter authors
have refined the idea in a number of important works, where they focused
on specific finite states signatures through an {\it instanton event
generator}. They also suggested dedicated experiments which were carried at
HERA. In these works the focus is on the mechanism in which DIS 
{\it resolves the instanton}, kicking 
the light quark out of its zero mode. The corresponding expressions
are exponentially suppressed at large momentum transfer, e.g. $e^{-Q\rho}$.
In our approach the quarks form a {\it frozen dipole}, in other words
they behave like  heavy static quarks and interact with instantons only
via its gluonic field. This mechanism is dominant at low $x$. Therefore,
our approach at low $x$ complements the approaches pursued at large $x$.

The closest in spirit to our treatment of DIS are recent works
by Schrempp and Utterman~\cite{SU}. In their most recent
paper (last citation) a brief reference to {\it Wilson loop
scattering} through one-instanton is made much along the lines we have advocated
in \cite{SZ1} for dipole-dipole scattering. Their starting
point is the optical theorem relating the total DIS cross
section to the imaginary part of the forward scattering amplitude.
However, they appear to use the one-instanton amplitude which
is purely {real}. The one-instanton cannot be {\it cut}~\cite{SZ1}.
Only rescattering, with an instanton in the amplitude and an
antiinstanton in the conjugate amplitude with explicit 
quasielastic or inelastic states crossing the unitarity cut,
lead to an imaginary part. Therefore, our results are overall
different. In particular, there is no {\it saturation} of the dipole cross
section  $\sigma(\dd)$ at large dipole size $\dd$.

This paper is one in line of a few we have written in the
past few years regarding high energy and low angle scattering through
instantons. A relation between instantons and diffractive physics (pomerons)
was suggested in~\cite{Shu_b_00}, where in particular
a relation between the DIS dipole cross section $\sigma(r)$ 
 were argued to be related to
the mean vacuum instanton size $\rho\approx 1/3$ fm~\cite{Shu_82}. 
The importance and the details on how to calculate the instanton-induced
quasi-elastic parton-parton and dipole-dipole scattering have been developed
in~\cite{SZ1}, with the resulting  small cross section
proportional to the instanton diluteness $squared$.
However, if one sums over all gluons radiated from the instanton
into the final state, which is equivalent
to the production of certain semiclassical
gluonic clusters -- the QCD sphalerons~\cite{OCS} -- the cross section
becomes much larger and is proportional to the
diluteness parameter in the {\it first} power.

As shown in \cite{NSZ,KKL},
this approach produces  reasonable parameters for the soft pomeron. 
A single sphaleron production gives the $\sigma_1\, {\rm ln}\,s$ term
in (\ref{eqn_crosssect}), and its t-channel iteration provides
higher powers of the ${\rm ln}\,s$. 
We have extended  such calculations to  double-pomeron processes and related
properties of QCD sphalerons with clusters and/or identified
hadrons produced in this way in \cite{SZ_doublepomeron}.
The relation between the so called ``color glass condensate'' and
sphalerons, especially in the early stages of
 heavy ion collisions, have been extensively discussed
in \cite{Shu_how}.

In this paper we  focus on the {\em double} DIS process,
 with two frozen small-size dipoles moving with a speed of
 light in opposite directions.
In this ultimate process, perhaps to be realized at the Next Linear Collider,
 there is no proton or other hadron in the process, and 
the non-perturbative glue which generates
the dipole-dipole scattering belongs to the 
 wave function of the QCD vacuum itself. 
One can think of it as zero point oscillations around
zero fields, resulting in the usual propagators,
and large-amplitude (semiclassically described) fields,
representing ``tails'' of the wave function under the barrier.
  We will not discuss here any of the 
 the vacuum instanton physics, which can be found e.g. in
review \cite{SS_98}. For our purposes it is enough to
remind the reader that the diluteness factor $\kappa=n_{\rm ins}\rho^4\approx 10^{-2}$
because the instanton density $n_{\rm ins}\approx 1/{\rm fm}^{4}$ and
the mean size of QCD instantons is $\rho\approx 1/3 \, {\rm fm}$.
The diluteness is formally $\sim {\rm exp}(-2\pi/\alpha_s(\rho))$,
which explains why they are invisible in perturbative diagrams.
This diluteness of the instanton vacuum leads to a relative diluteness
of the semiclassical glue in the proton, and is behind 
the small shadowing corrections.

In this paper we
extend our earlier results to inelastic dipole-dipole scattering, 
the basic process of great interest to low $x$ DIS.
We will show that modulo the enhancement in the overall cross section
caused by the production of a sphaleron, our inelastic 
cross section is in agreement with the quasi-elastic instanton amplitude for
dipole-dipole scattering~\cite{SZ1}. In particular, the inelastic 
cross section scales with the squared dipole size for small dipoles, and asymptotes
the squared sphaleron (instanton) size for large dipoles.

Even without a calculation, the instanton approach explains why the
field is strong and classical, $\approx 1/g$, and why it appears in small-size spots.
Such qualitative  discussion was in the literature for decades,
see e.g.~\cite{Shu_822}. The diluteness of the
instanton vacuum insures that although there are many partons (dipoles) in a
nucleon, presumably only about one may  meet the tunneling process  
active at the moment and near the place
of the collision. Phenomenologically, a simple qualitative
eikonal estimate in $pp$ and $p\bar{p}$ shows that

\be
\frac{\sigma_{\rm el}}{\sigma_{\rm in}}\approx \frac{1-\sqrt{\bf
p}}{1-{\bf p}}\approx \frac 14
\nonumber
\ee
at $\sqrt{s}\approx 30 \,{\rm GeV}$. For a Poisson distribution the
probability of no collision ${\bf p}\approx e^{-\left<S\right>}$ where 
$\left<S\right>\approx 0.8$ is the average number of inelastic pair
collisions. This is suggestive of less than one sphaleron per collision
at such energies produced. For DIS, with its single small dipole,
it can  only be a single instanton involved. 

The inelastic dipole-dipole cross section has the generic structure, 
\be
\sigma_{IN}\approx {\rm ln}\,s\,\,{\bf d}^2\overline{{\bf d}}^2\rho^2
n_{\rm ins} 
\nonumber
\ee
times a dimensionless function of $\dd/\rho$ and $\overline{\dd}/\rho$
as well as the impact parameter ${\bf b}/\rho$, to be discussed below.
This scaling has already been noted in~\cite{SZ1} for 
quasi-elastic processes. Again, the
logarithm of the energy comes from the longitudinal phase space
of the sphaleron. We recall that the cross section is
of {\it first} order in the instanton density, not second order
and vanishes whenever any of the parameters
${\bf d},\overline{{\bf d}},\rho$ goes to zero.

As discussed in \cite{NSZ,KKL} and elsewhere, the semiclassical
sphaleron production cross section is not yet calculated up to 
an absolute numerical constant. Therefore we focus on the predicted dependence
on the variables in question, the dipole sizes and the impact parameter.
Strictly speaking, the dipole-dipole cross section should be used
 for the double DIS $\gamma^*\gamma^*$ reactions, which however has
 not yet been studied in  sufficient details.
We show however, that the dependence on the dipole size and especially on the
impact parameter is in agreement with current HERA data on   $\gamma^*
p$, if we were to think of a nucleon as a color dipole as well.

\begin{figure}[h]
\centering
\includegraphics[width=6cm]{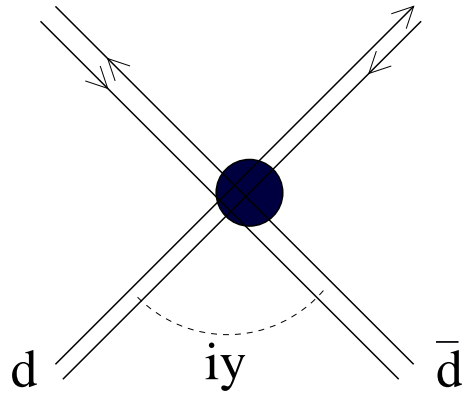}
\vskip 0.5cm
\caption{\label{fig_dipdip} 
Two dipoles of sizes $\dd,\dbar$ 
propagate by straight crossing lines, with the Euclidean angle between
them being $iy$ where $y$ is the relative rapidity. The centers
of these dipoles  are also separated by a minimal impact parameter $\bb$
(not shown). The dashed circle
near the collision point is the instanton, its radius is $\rho$.
}
\end{figure}

\section{Dipole-Dipole Scattering}

In this section we derive the general form of the inclusive
dipole-dipole scattering amplitude using QCD sphalerons in
for large $\sqrt{s}$ as discussed in \cite{SZ1,NSZ}. We will
briefly recall the dipole-dipole structure through pertinent
Wilson lines, and then derive the general form of the 
inelastic cross section which is our main result. The
quasi-elastic cross section with only color-flip in the 
intermediate state follows by inspection.

\subsection{Eikonalized Dipole-Dipole}

The notations we use and the overall setting of the calculation is
explained
in Fig.\ref{fig_dipdip}. Each dipole is represented
by a pair of Wilson lines moving by straight lines. The
scattering amplitude is proportional to the vacuum
expectation value of these 4 Wilson lines. Expansion in the field
strength
$g\,A_\mu$ would generate the pQCD diagrams. However for instanton
fields the expansion is not possible and each quark 
passing through the instanton has a color rotation of the
order one, see~\cite{SZ1} for details.

 Let $AA$ be the initial color of the dipole
and $CD$ its final color. The Wilson loop with open
 final color  for the dipole 
configuration in the one-instanton background is~\cite{SZ1}
\be
&&{\cal W}_{AA}^{CD} (\theta , b) =
{\rm cos}\,\alpha_-\, {\rm cos}\, \alpha_+\, {\bf 1}_{CD}\nonumber\\
&&+i{\rm cos}\,\alpha_-\,{\rm sin}\,\alpha_+\,{\bf R}^{ab}\,\hat{\bf n}_+^b\,(\tau^a)_{DC}\nonumber\\
&&-i{\rm sin}\,\alpha_-\,{\rm cos}\,\alpha_+\,{\bf R}^{ab}\,\hat{\bf n}_-^b\,(\tau^a)_{DC}\nonumber\\
&&+{\rm sin}\,\alpha_-\,{\rm sin}\,\alpha_+\,{\bf R}^{ab}\,{\bf R}^{cd}\hat{\bf n}_-^b\, \hat{\bf n}_+^d\,
(\tau^c\tau^a)_{DC}\,\,.
\label{DDXX1}
\ee
We have defined 
\begin{eqnarray}
\alpha_\pm =&& \pi\left( 1-\frac {\gamma_\pm}{\sqrt{\gamma_\pm +\rho^2}}\right)\nonumber\\
\gamma_\pm^2 =&& (y_4{\rm sin\theta} - y_3 {\rm cos\theta} )^2 + ({\bf
y}\pm {\bf d}/2)^2\nonumber\\
{\bf n}_+\cdot {\bf  n}_- =&&
(y_4\,{\rm sin}\theta - y_3\, {\rm cos}\theta )^2 + 
{\bf y}^2 -{{\bf d}^2}/4 
\label{DDI5}
\end{eqnarray}
with ${\bf n}_\pm\cdot{\bf n}_\pm=\gamma_\pm^2$. 
The dipole is sloped at an angle $\theta$ from 
the $y_4$ direction, but otherwise is at an arbitrary location away from
the instanton centered at the origin.

The dipole-dipole scattering amplitude 
with open-color in the final state can be
constructed by using two dipole configurations
as given by (\ref{DDXX1}) with a relative angle $\theta$.
After averaging over the instanton color-orientations, the
dipole-dipole amplitude is~\cite{SZ1}

\be
&&{\cal W}_{AA}^{CD} (\theta, b)\,{\underline{\cal W}}_{A'A'}^{C'D'} (0,0) =\nonumber\\&&
\frac 2{N_c}\,{\bf W}_1\,\,{\bf 1}_{CD}\,{\bf 1}_{C'D'} +\frac 1{N_c^2-1}\,
{\bf W}\,\,(\tau^a)_{DC}\,(\tau^a)_{D'C'}\,\,.\nonumber\\
\label{DDOPEN}
\ee
The singlet part ${\bf W}_1$ drops after analytical continuation at
large $\sqrt{s}$ and will be ignored. The non-vanishing part of the 
dipole-dipole cross section is carried by the (octet) part 

\be
{\bf W}\approx &&
-{\rm cos}\,\alpha_-\,{\rm sin}\,\alpha_+
{\rm cos}\,\underline\alpha_-\,{\rm sin}\,\underline\alpha_+ 
{\bf n}_+\cdot\underline{\bf n}_+\nonumber\\&& 
-{\rm sin}\,\alpha_-\,{\rm cos}\,\alpha_+
{\rm sin}\,\underline\alpha_-\,{\rm cos}\,\underline\alpha_+ 
{\bf n}_-\cdot\underline{\bf n}_-\nonumber\\&& 
+{\rm cos}\,\alpha_-\,{\rm sin}\,\alpha_+
{\rm sin}\,\underline\alpha_-\,{\rm cos}\,\underline\alpha_+ 
{\bf n}_+\cdot\underline{\bf n}_-\nonumber\\&& 
+{\rm sin}\,\alpha_-\,{\rm cos}\,\alpha_+
{\rm cos}\,\underline\alpha_-\,{\rm sin}\,\underline\alpha_+ 
{\bf n}_-\cdot\underline{\bf n}_+\,\,,
\label{DDOCTET}
\ee
where the underlying notation refers to a dipole of size $\overline{{\bf d}}$
at angle $\theta=0$ with respect to the $x_4$ axis, but again
at an arbitrary location away from the instanton located at the 
origin.

The specificity of (\ref{DDOCTET}) is that for a fixed size dipole
$\overline{{\bf d}}$, as the varying size dipole increases, say
$|{\bf d}|\gg |\overline{{\bf d}}|$ with $\alpha_-\approx 0$, it reduces to

\be
{\bf W}\approx &&
-{\rm sin}\,\alpha_+
{\rm cos}\,\underline\alpha_-\,{\rm sin}\,\underline\alpha_+ 
{\bf n}_+\cdot\underline{\bf n}_+\nonumber\\&& 
+{\rm sin}\,\alpha_+
{\rm sin}\,\underline\alpha_-\,{\rm cos}\,\underline\alpha_+ 
{\bf n}_+\cdot\underline{\bf n}_-
\label{DSIMPLE}
\ee
which is the scattering of a point charge on a fixed size
dipole $\overline{{\bf d}}$. This is a property of the unexpanded
sine and cosine contributions (re-summed gluon lines) stemming
from the unitary character of the Wilson lines. It will prove
important below for our discussion on asymptotia.

\subsection{Inelastic Dipole-Dipole Cross Section}

In terms of (\ref{DDOCTET}) the inelastic cross section 
for dipole-dipole scattering through sphalerons follow the 
general construction developed in \cite{NSZ} for parton-parton
(charge-charge) scattering. To leading logarithm accuracy, the
result in Minkowski space is

\be
&&\sigma_{IN} (s) \approx \,\,{\rm ln s}\,\,{\rm Im}\,\sum_{CD,\underline{C}\underline{D}} 
\,\frac{1}{(2\pi)^6}\,\int dQ^2 \, dq_{1\perp} \, dq_{2\perp }\,\nonumber\\
&&\times\int\,dz\,d\dot{I}\,d\dot{I}'\,e^{iQz + iS(\dot{I})-
iS(\dot{I}')+ iS(\dot{I}, \dot{I}',z)}\nonumber\\
&&\times\int dx_-dx_\perp dy_+dy_\perp \,e^{-iq_{1\perp}
x_\perp -iq_{2\perp}y_\perp}\nonumber\\
&&\times {\cal W}_{AA}^{CD}(\infty, x_-, x_\perp) 
\underline{\cal W}_{\underline{A}\underline{A}}^{\underline{C}\underline{D}}
(y_+, \infty, y_\perp) 
\nonumber\\
&&\times\int dx'_-dx'_\perp dy'_+dy'_\perp \,e^{-iq_{1\perp}
x'_\perp -iq_{2\perp}y'_\perp}\nonumber\\
&&\times {\cal W}_{AA}^{CD\,*} (\infty, x'_-, x'_\perp) 
\underline{\cal W}_{\underline{A}\underline{A}}^{\underline{C}\underline{D}\,*}
(y'_+, \infty, y'_\perp) 
\,\,,
\nonumber\\
\label{IN1}
\ee
where the collective integrations $z,\dot{I},\dot{I}'$ correspond to
localized chromomagnetic field in space-time. The appearance of ${\rm ln}\,s$
underlines the fact that the integrand in (\ref{IN1}) involves only 
$Q^2$ which is the transferred mass in the inelastic half of the
forward amplitude, and $q_{1,2\perp}$ which are the transverse 
transferred momenta through the dipole form-factors. 

We now elect to evaluate (\ref{IN1}) in Euclidean space by interpreting
the localized fields as singular gauge fields (sphaleron) that maximize 
the partial scattering amplitude in $Q^2$, and the external dipole
amplitudes with (\ref{DDOPEN}-\ref{DDOCTET}) with the identification
$\theta\leftarrow i{\rm ln}\,s$. The result is

\be
{\sigma_{IN}} (s)\approx \int \,\frac{d{\bf q}}{(2\pi)^2}\,
\left|{\int\,d{\bf b}}\,e^{-i{\bf q}\cdot{\bf b}}\,{\bf F}(s,{\bf b})\right|^2
\label{IN2}
\ee
where 
\be 
&&{\bf F}(s,{\bf b}) =i\,\sqrt{{\kappa\,{\rm ln}\,s}}\,\frac{4}{(N_c^2-1)}
\frac {{\bf d}\cdot{\overline{{\bf d}}}}{8\pi\rho^2}\,{\bf I}(\dd/\rho,\dbar/\rho,\bb/\rho)
\label{eqn_F_def} \ee
defined via the dimensionless function of the three 2-d vector variables 
\be \label{eqn_I_def} 
&& {\bf I}(\dd/\rho,\dbar/\rho,\bb/\rho)\equiv{1\over \rho^2}\int\,d{\bf{R}}\,dx_-\,dy_+\nonumber\\
&&\times
\left(\frac{c_-s_+\underline{c}_-\underline{s}_+}{\gamma_+\underline{\gamma}_+}
+\frac{s_-c_+\underline{s}_-\underline{c}_+}{\gamma_-\underline{\gamma}_-}
+\frac{c_-s_+\underline{s}_-\underline{c}_+}{\gamma_+\underline{\gamma}_-}
+\frac{s_-c_+\underline{c}_-\underline{s}_+}{\gamma_-\underline{\gamma}_+}\right)
\nonumber\\
\label{IN2x}
\ee
after $({\bf b}\rightarrow -{\bf b}, {\bf d}\rightarrow \overline{\bf d})$ symmetrization,
and rotation back to Minkowski space in the large $\sqrt{s}$ limit. 
The sine $s_\pm$ and cosine
$c_\pm$ functions in (\ref{IN2x}) are defined with arguments
$\alpha_\pm$, and similarly for $\underline{\alpha}$ with the substitution
$\gamma\rightarrow\underline{\gamma}$, where

\be
\gamma_\pm^2=&&x_-^2+({\bf R}+{\bf b}/2\pm {\bf d}/2)^2\nonumber\\
\underline{\gamma}_\pm^2=&&y_+^2+({\bf R}-{\bf b}/2\pm \underline{{\bf d}}/2)^2\,\,.
\label{IN4}
\ee
The contributions arising from (\ref{IN2x}) is mostly through
the {\it electric} dipoles, and vanish for zero size
dipoles.  The magnitic contribution is of the form
$\dd\times\overline{\dd}$ and drops under symmetrization.
The quasi-elastic cross section with only color-flip
in the intermediate states is due to the instanton-antiinstanton
as opposed to the sphaleron configuration used here. 
It is readily checked that it follows from (\ref{IN2}) through
the substitution $\kappa\,{\rm ln}\,s\rightarrow \kappa^2$.

\subsection{Eikonal Unitarization}

The inelastic cross section (\ref{IN2}) can be readily rewritten
as

\be
{\sigma_{IN}} (s)\approx \int \,{d{\bf b}}\,\left|{\bf F}(s,{\bf b})\right|^2
\label{EU1}
\ee
which allows $|{\bf F}(s,{\bf b}|^2$ to be interpreted as the absorption
probability at fixed impact parameter and energy
\footnote{With the addition of the constant part
as in (\ref{00}).}. In the present
case, the absorption probability grows like ${\rm ln}\,s$ with
large $\sqrt{s}$ and therefore upsets unitarity constraints (albeit
logarithmically). A simple way to fix this is by using the standard
eikonal (s-channel) unitarization, i.e.

\be
|{\bf F}(s,{\bf b})|^2 \rightarrow 1-e^{-|{\bf F}(s,{\bf b})|^2 }
\label{EU2}
\ee
which yields the eikonalized cross sections

\be
&&\sigma_{TOT}(s)\approx 2\int\,{d{\bf b}}\,
\left(1-e^{-\frac 12 |{\bf F}(s,{\bf b})|^2 }\right)\nonumber\\
&&\sigma_{EL}(s)\approx \int\,{d{\bf b}}\,
\left|1-e^{-\frac 12 |{\bf F}(s,{\bf b})|^2 }\right|^2\nonumber\\
&&\sigma_{IN}(s)\approx \int\,{d{\bf b}}\,
\left(1-e^{-|{\bf F}(s,{\bf b})|^2 }\right)\,\,.
\label{EU3}
\ee
The total cross section in (\ref{EU3}) saturates when the distance
between the dipoles (impact parameter) exceeds the sphaleron size
$\rho$.

\subsection{The dependence on the dipole size and impact parameter}


If the dipoles sizes are small, one expects the dipole
approximation to hold and the cross section to be proportional
to $\dd^2$ or  $\dbar^2$. Such factors are already included
in the definition of the function ${\bf F}$ (\ref{eqn_F_def}), and thus
one expects a smooth limit of ${\bf I}(\dd/\rho,\dbar/\rho,\bb/\rho)$
when either of the dipoles, or both go to zero. A look at its
definition confirms that it is indeed the case.

Now let us qualitatively discuss the \bb-dependence. One naturally
expects a maximum at \bb=0, so at small impact parameters 
${\bf I} (\bb)={\bf I}(\bb=0)(1-{\rm const}\,
\bb^2/\rho^2)$. At large $\bb$ one expects a power decrease. Indeed, if
$\bb\gg \rho$ one may think that the main contribution to the cross section
comes from the instanton far from each of the lines. In such a case
all $\gamma_\pm\approx {\bf b} \gg \rho$ are large compared to
$\rho$ and the arguments of all
 trigonometric functions
in (\ref{eqn_I_def}) are small $\alpha_\pm\approx\rho^2/2\gamma_\pm^2$, so that
all cosines and sines can be replaced by either 1 or their arguments.
Physically this corresponds to a single-gluon exchange
between each dipole and the instanton.
As a result, our 4-dimensional integral over $x_-,y_+,R_x,R_y\approx \bb$
will have an integrand $\approx 1/\gamma^6$ and is therefore convergent,
producing ${\bf I}(0,0,\bb/\rho)\sim  \rho^2/\bb^2$ 
with a positive coefficient. 

Let us now discuss the issue quantitatively. In our preceding papers,
see for instance \cite{SZ_doublepomeron}, we have argued that (unlike many other
instanton applications) the modification of the
instanton profile is numerically important for high energy collisions.
The dipole forces at large distances are of course true for massless
gluons only and are ultimatly absent in confining theories. A
phenomenological way to enforce this was discussed in
\cite{SZ_doublepomeron}. In Fig. \ref{fig_shapefit} we show how two selections
of cutoffs on the range of the instanton field, with

\be f_1=e^{-0.5\,\gamma} \hspace{1cm} f_2=e^{-0.08\, \gamma^2} 
\label{eqn_shapemod} 
\ee
correspond to the cross section of double-pomeron pp events from WA102
experiment at CERN. The modification function enters all the arguments
of the trigonometric function in ${\bf I}$ (\ref{eqn_I_def}) through
the substution $\alpha\rightarrow \alpha\,f$  with the
pertinent $\gamma$ arguments. Note that the instanton size is kept fixed at
$\rho=1/3$ fm. The inclusion of a realistic size distribution would somewhat
change the small and large $t$ tails.

\vskip .5cm
\begin{figure}[t!]
\centering
\includegraphics[width=7cm]{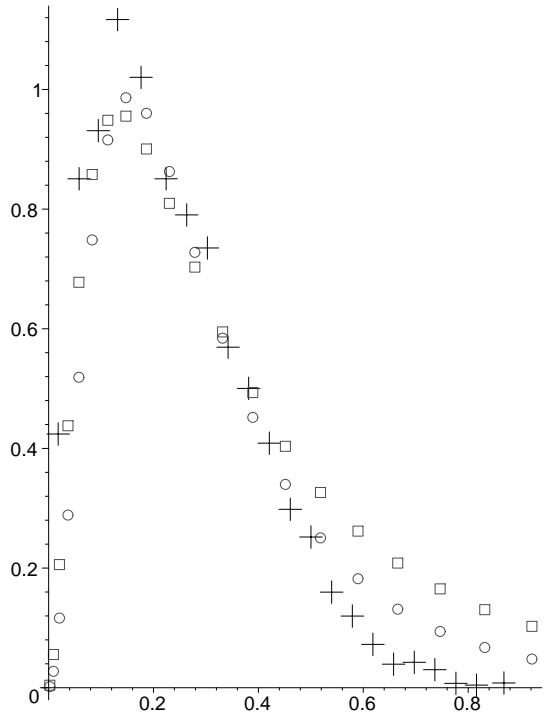}
\vskip 0.5cm
\caption{\label{fig_shapefit} 
Instanton shape modification compared to WA102 experimental data (crosses,
no error bar shown) on the momentum transfer $t$ (the same for both
protons). The squares and circles correspond to the
instanton form-factor to the fourth power, with the cutoffs
$f_{1,2}$ (\ref{eqn_shapemod}), respectively. The cutoffs
produce the dip at small $t$ seen in the data.}
\end{figure}

\vskip 1cm
\begin{figure}[t]
\centering
\includegraphics[width=8cm]{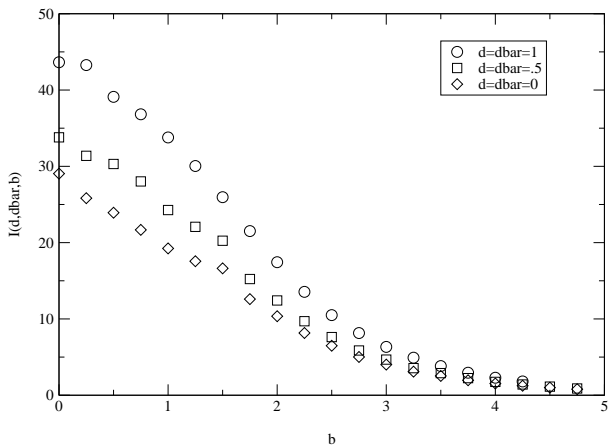}
\vskip 0.5cm
\caption{\label{fig_alld} 
Dependence of the function ${\bf I}(\dd/\rho,\dbar/\rho,\bb/\rho)$ 
defined in (\ref{eqn_I_def}) on its last argument, the impact
parameter \bb.  The case shown is for all 3 vectors in the same
plane and for equal dipole sizes (in units of the instanton size).
}
\end{figure}

Now we turn to a more quantitative discussion of (\ref{eqn_I_def}) with the
cutoffs (\ref{eqn_shapemod}). In Fig.~\ref{fig_alld}  we show the
$\bb$-dependence of the pertinent integral, for few values of the dipole sizes.
Note also, that if the dipoles are of different sizes, the largest
completely dominates. The calculations were carried out using
Monte-Carlo with few checks through Mathematica.

In Fig.~\ref{fig_d_dep} we show the value of this function
when \bb=0 and both dipoles with equal size.
There is almost no dependence on $\dd, \dbar$ when the dipoles  
are less than about $\rho/2$. The dependence appears when the dipole sizes are comparable
to $\rho$, but desappears again 
when $\dd, \dbar\gg \rho$. Remarkably, a much weaker dipole size dependence
is found when only one of the dipole size is varied while the other is
fixed. For example, if $\dbar=\rho$ the dependence on $\dd$ is of the
order of few percents in the entire range of the
dipole size variation, from small to large.

Finally, we note that 
the numerical value of the dimensionless function ${\bf I}$ is not of the
order of 1 but of the order of $20$, which is large enough to compensate for a
square root of the instanton diluteness. This shows that in a
relevant volume near the dipole collision point there is about 1 instanton
in the QCD vacuum, and so the cross section is not really suppressed.

\vskip .5cm
\begin{figure}[t!]
\centering
\includegraphics[width=7cm]{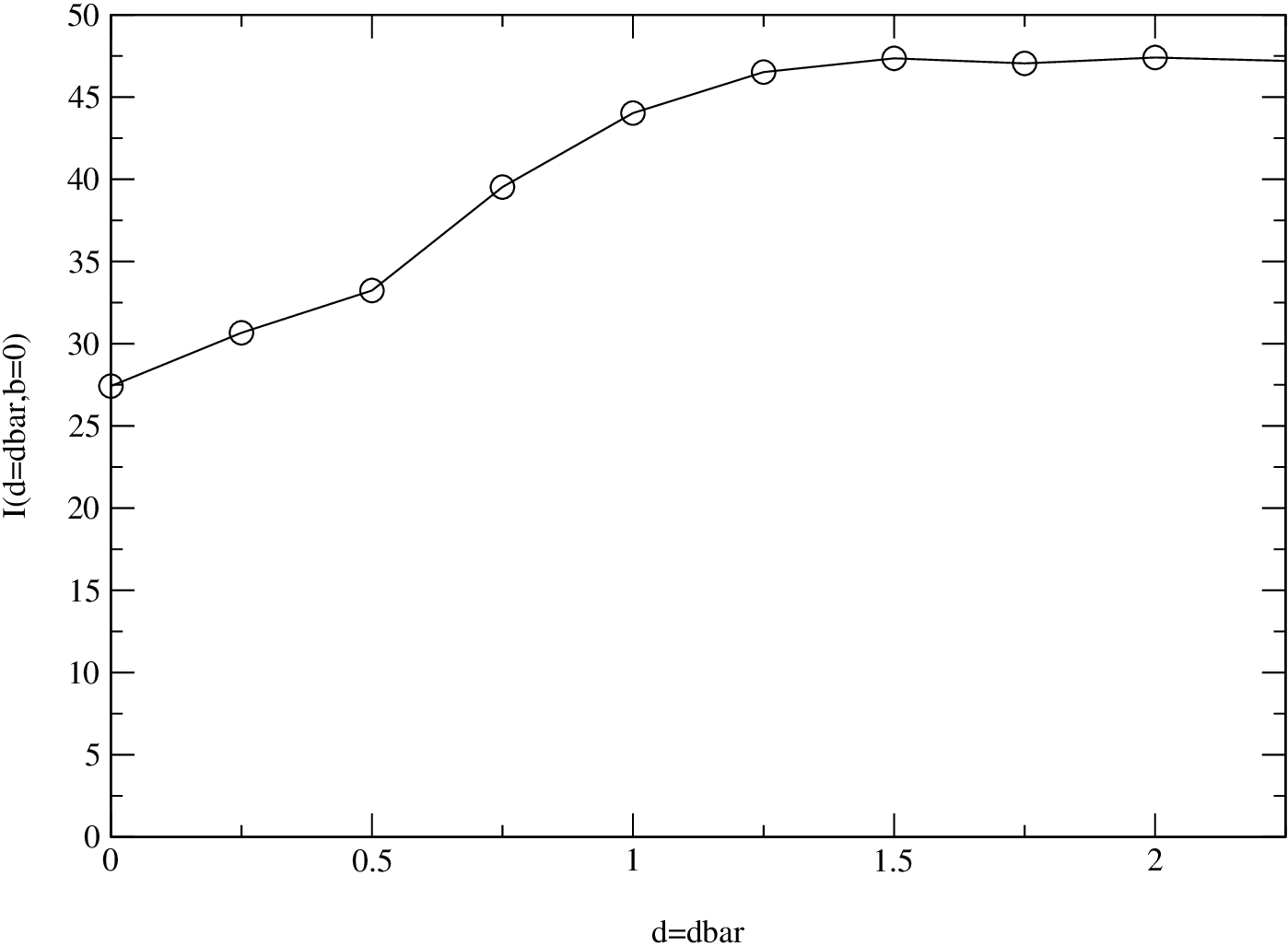}
\vskip 0.5cm
\caption{\label{fig_d_dep} 
Dependence of the function ${\bf I}(\dd/\rho,\dbar/\rho,\bb/\rho)$ 
defined in (\ref{eqn_I_def}) at $\bb=0$ on the dipole size 
$\dd/\rho=\dbar/\rho$.
}
\end{figure}

\section{Dipole-dipole cross section and DIS phenomenology}

  The main point of this paper is that the glue
on which DIS is made is basically from the vacuum,
one does not need a nucleon or other quark states.
In principle, one can directly compare our results with the
experimental data for $\gamma^* \gamma^* $ collisions. 
We know rather well how a photon becomes a
frozen $\bar q q$ dipole. It is well described by a simple
loop diagram\footnote{By the way,
the vector current is a special case,
in which all non-perturbative corrections tend to cancel to a high
accuracy, as studies of correlations functions deduced from $e^+e^-$
and $\tau$ decays indicate, see \cite{Shu_822}. This is not the case 
for scalar or pseudoscalar ``photons'', if Nature were to provide
them.}, see e.g. \cite{fms}. The size of a small dipole is related to $Q^2$ by
 $d\approx \pi/Q$.

For sufficiently large $Q^2$ and $\overline{Q}^2$ the dipole sizes 
are small compared to $\rho$ and one may set \footnote{
We recall that the large-$Q$ or small-${\bf d}$ dependence
is given by DGLAP evolution.  We are only
discussing the nonperturbative aspects of the problem,
ignoring radiative corrections. Our results
 should be treated as the
``initial conditions'' for DGLAP at the semi-hard scale $Q\approx 1\, {\rm
GeV}$.}  $\dd=\dbar=0$ in our function ${\bf I}$
(the lowest set of points in Fig.~\ref{fig_alld}). Note that
 the impact parameter profile  is quite narrow,
with a width of about the instanton size. 

Unfortunately, past LEP experiments had very small acceptance for such
events and had very small statistics. The diffractive events, such as
 $\gamma^*p\rightarrow J/\psi\,X$, have not been yet studied well enough
 to test our predictions. 
There are of course much
 better  DIS data on a near-real photon,  $\gamma^* \gamma $, 
from LEP experiments. Those correspond to the asymmetric
case with one small and one large dipole, which is close to
the upper set of points in Fig.~\ref{fig_alld}. 

The best DIS data are from HERA, for the scattering on a proton  $\gamma^* p $.
The observed dependence of $\sigma_{\gamma*p}$ on ${\bf d}$ is a matter of 
ongoing debate. The original model by Golec-Biernat and Wusthoff
\cite{GBW} assumes that it grows as $\sigma_{\gamma*p}\approx {\bf d}^2$ at small
${\bf d}$ and then saturates, at some constant cross section
$\sigma_0\approx 20$ mb.  Alternative parameterizations of the dipole
cross section were also considered, e.g. in~\cite{fms}. 
A simultaneous fit to DIS and diffractive
$\gamma^*p\rightarrow J/\psi+...$ in~\cite{TK} resulted in different conclusions:
({\bf i}) there is more or less quadratic dependence on the dipole size
$\sigma_{\gamma*p}\approx {\bf d}^2$ for $all$ sizes
with no need for saturation or unitarization; 
({\bf ii}) The $\bb$-dependence is given by a function $T({\bf b})$ of 
roughly Gaussian shape with an rms radius $\bb_{\rm rms}=0.4-0.5
\,{\rm fm}$.

Strictly speaking we do not know if a proton can be seen as a
dipole (or a distribution of  dipoles). One may still try to use
our calculation, thinking about the 
 the nucleon as a single large dipole ($\dbar\approx \rho$). 
If so, our results qualitatively  agree with Kowalski and Teaney recent
observations as mentioned above. In particular, the
dependence on the dipole sizes is quadratic without
saturation and the  width
of the ``gluonic spot'' we predict turns out to be also about 1/2 fm.
We plan to  do a more specific model analysis for the proton and carry a
detailed discussion of DIS and diffractive DIS in a separate publication.

\section{Acknowledgments}

This work was supported in parts by the US-DOE grant DE-FG-88ER40388.
We thank H. Kowalski, D. Teaney and G. Papp for valuable discussions, and 
A. Mueller for attracting our attention to the impact parameter issue in 
diffractive DIS. We also thank E. Levin, M. Lublinsky and
F. Schrempp for helpful correspondence.

\end{narrowtext}
\end{document}